\documentstyle[12pt]{article}

\begin{document}

\title{Diagonal quantum  Bianchi type IX models in $N = 1$ supergravity}

\author{A.D.Y. Cheng\thanks{email: adyc100@cus.cam.ac.uk} and  P.D. D'Eath\thanks{
email: pdd1000@damtp.cam.ac.uk} \\ {}\\ {\sf DAMTP} \\ University of Cambridge \\ Silver Street \\
Cambridge CB3 9EW\\United Kingdom}

\date{}
\maketitle

\begin{abstract}
We take the general quantum constraints of $N=1$ supergravity in the special case of a 
Bianchi metric, with gravitino fields constant in the invariant basis.
We construct the most general possible wave function which solves the Lorentz constraints 
and study the supersymmetry constraints in the Bianchi Class A Models.
For the Bianchi-IX cases, both the Hartle-Hawking state
and wormhole state are found to exist in the middle fermion levels.
\end{abstract}

\pagebreak

\section{\bf Introduction}

Since the discovery of supersymmetry 20 years ago, many people have been fascinated by supergravity theories. There
are several reasons for this. First, supergravity theories are the only consistent theories which couple
fundamental spin-3/2 particles to gravity. Second, supergravity theories are less divergent than general 
relativity. There are some indications that pure $N=1$ supergravity is finite \cite{finite}.

The canonical formulation of $N=1$ supergravity was presented in ref.\ \cite{pilati} in four-component spinor 
notation and in ref.\ \cite{84} in two-component spinor notation. In finding a physical state, it is sufficient 
to solve the Lorentz and supersymmetry constraints of the theory; the algebra of constraints implies that physical 
wave function will also obey the Hamiltonian constraints \cite{84}.

In the past ten years, there has been active research in supersymmetric quantum cosmology, especially in 
$N=1$ supergravity theory. Bianchi class A models of pure $N=1$ supergravity were studied in 
refs.\ \cite{P, P1, P2, gl} using both triad and Ashtekar variables. These authors assumed a simple Ansatz for the 
wave function in the investigation of supersymmetric quantum cosmology. They found that only simple solutions  
were present in the bosonic and full fermionic sectors of the wave function. This curious result was
joined by yet another disturbing one. When a cosmological constant was added, it appeared that there was no 
non-trivial physical wave function \cite{PP}. One might think that supersymmetric quantum cosmology is not very
interesting. 

However, recently, Csord\'as and Graham \cite{cg} pointed out there exist middle fermion
states in the minisuperspace models of pure $N = 1$ supergravity. They
showed that there is a richer structure of physical states  of supersymmetric
quantum cosmology than that found in previous works \cite{P,P1,P2,gl,PP}. They rightly criticise  the
Ansatz for the wave function used in refs.\ \cite{P,P1,P2,gl,PP} as not being general enough.
One now needs to investigate these middle states in the fullest possible detail.

The wave function of the universe of supersymmetric quantum cosmology can be expanded in even numbers of 
gravitinos up to order 6. Since we have 6
gravitinos, there are  ${6 \choose 2} = 15$ allowed terms of two fermions.
In this sense, the Ansatz used in  refs.\ \cite{P,P1,P2,gl,PP}, which has only two degrees 
of freedom at the two-fermion level, is not general enough and this is the reason why
 refs.\ \cite{P,P1,P2,gl,PP} failed to find the interesting middle fermion states.
Csord\'as and Graham  \cite{cg} constructed a new Ansatz for the wave function
based on a scalar function  $f(h_{pq})$, where $h_{pq}$ is the three-metric of the space-like hypersurface. 
For the two-fermion level, they noticed first that 
$\bar S_{A'} \bar S^{A'} f(h_{pq})$ automatically solves the Lorentz constraints,
where $\bar S_{A'}$ is the supersymmetry constraint operator. 

They further noticed that this expression
 solves the $\bar S_{A'}$ supersymmetry constraint, using the anti-commutation properties. The only 
constraint that remains to solve is the $S_{A}$ constraint. By solving this constraint, they reduce the
problem to solving the Wheeler-DeWitt equation for $f$.
This approach is, however, limited by being based on an Ansatz. Further information can be 
obtained by studying the complete set of coupled first-order partial differential  constraint equations, 
as is done here.

 We start from the wave function which is the most general solution to the Lorentz constraint.\footnote{The Ansatz 
$\bar S_{A'} \bar S^{A'} f(h_{pq})$ only works when there are no chiral breaking terms in the 
supersymmetry constraints (see section 4).}
In section 2, we will briefly describe the conventions and variables to be used in 
the calculations. We will carry out the dimensional reduction from $ 3 + 1 $ to $ 0 + 1$ dimensions.
From the reduced action, the supersymmetry constraints are found. It is sufficient,
in finding a physical state, to solve
the Lorentz and supersymmetry constraints of the theory \cite{84,T}. Because of
the anti-commutation relations $\left[ S_A,~ \bar S_{A'} \right]_+ \sim {\cal H}_{A A'} $,
the supersymmetry constraints $ S_A \Psi = 0,~ S_{A'} \Psi = 0 $ on a
physical wave function $ \Psi $ imply the Hamiltonian constraints $ {\cal H}_{A A'}
\Psi = 0 $ \cite{84,T}. We study the supersymmetry constraints
which are  a set of coupled first-order partial differential equations for the
components of the wave function. 
We find, for the case of a diagonal Bianchi IX model, that of the 15 possible coefficients at 
two-fermion level, the coefficients of 9 are zero. Only the remaining 6 are dynamical.
In section 3, we will make a comparision with the work of 
Csord\'as and Graham \cite{cg}. Section 4 contains the conclusion.

\section{\bf Dimensional Reduction and Derivation of the Supersymmetry Constraints}

Using two-component spinors \cite{84}, the action \cite{WB} is
\begin{equation}
S = \int d^4 x \left[ {1 \over 2} \left( \det e \right) R + {1 \over 2} 
\epsilon^{\mu \nu \rho \sigma} \left( \psi^{~A'}_{\mu} e_{\nu A A'}
D_\rho \psi^{~A}_{\sigma} + H.c. \right) \right]~.
\end{equation}
Here the tetrad is $ e^{~a}_{\mu} $ or equivalently $ e^{~A A'}_{\mu} $. The
gravitino field $ \left( \psi^{~A}_{\mu}, \psi^{~A'}_{\mu} \right) $ is
an odd (anti-commuting) Grassmann quantity. The scalar
curvature $ R $ and the covariant derivative $ D_\rho $ include torsion.

For the Bianchi class A models, we take the usual homogeneity conditions for the
Ansatz of the triads and the spatial gravitino fields. This  means that when the triad
$e_{p}^{~a}$ and the spatial gravitino $\psi_{p}^{~A}$ field are expanded with
respect to the invariant basis 
of the spatial hypersurface, the components are functions of time only. The $p$ indices are
the invariant indices which take the values $1,2,3$. We also assume that the time component
of the gravitino field $\psi_{0}^{~A}$ is a function of time only. One applies these homogeneity 
conditions to the above Lagrangian and carries out the 3-dimensional integration over the hypersurface
and then performs the Legendre transformations. The classical supersymmetry constraints are 
found to be:
\begin{equation}
\bar S_{A'} = \epsilon^{pqr} e_{p A A'}~\omega_{q~B}^{~A} \, \psi_{r}^{~B} - {1 \over 2} \, i 
 \, \psi_{p}^{~A} \, p^{p}_{~A A'}
\end{equation}
and the conjugate $ S_A $. Here $ n^{A A'} $ is the spinor version of the
unit future-pointing normal $ n^\mu $ to the surface $ t = {\rm const} $. It
is a function of the $ e^{~A A'}_{p} $, defined by
\begin{equation}
 n^{A A'} e_{p A A' } = 0~, \ \ \ \ n^{A A'} n_{A A'} = 1
\end{equation}
where $\omega_{pAB}$ is the torison free connection of spatial hypersurface. There is as usual 
a pair of second class constraints between the $\psi$ , $\bar \psi$ and their conjugate
momenta. We have to introduce the Dirac bracket to get rid of this pair
of second class constraints. With the help of ref.\ \cite{84}, we have the following 
bracket relations after the elimination:
\begin{eqnarray}
\left[ e_{p}^{~AA'} , p^{q}_{~BB'} \right]^* & =  & \delta_{A}^{B} \, \delta_{A'}^{B'}  \, \delta_{p}^{q}
\nonumber \\ \left[ \psi_{p}^{~A} , \bar \psi_{q}^{~A'} \right]^*_+ & = & - D_{pq}^{AA'}
\end{eqnarray}
where $D_{pq}^{AA'} = - 2i e_{q}^{~AB'} e_{p BB'} n^{BA'}$. The
rest of the brackets are zero.

Quantum mechanically, one replaces Dirac brackets by anti-commutators if both
arguments are odd or commutators if otherwise:
\begin{displaymath}
\left[E_1,E_2\right] = i\hbar \left[E_1,E_2\right]^*~, ~\left[O,E\right] = i \hbar \left[O,E\right]^*
~,~\left\{O_1,O_2\right\} =i \hbar \left[O_1,O_2\right]_+^*~. 
\end{displaymath}
We choose $ e_{p}^{~AA'}$ and $\psi_{p}^{~A}$ as our position coordinates and $p^{q}_{~AA'}$
and $\bar \psi_{q}^{~A'}$ as our momentum operators:
\begin{displaymath}
p^{q}_{~AA'} \rightarrow - i\hbar {\partial \over \partial e_{q}^{~AA'}}~~ 
, ~~\bar \psi_{q}^{~A'} \rightarrow -i\hbar D_{pq}^{AA'} {\partial \over \partial \psi_{p}^{~A}} 
\end{displaymath}
In the general theory, the corresponding quantum constraints read:
\begin{equation}
\bar S_{A'} = \epsilon^{pqr} \, e_{pAA'} \, D_q \psi_r^{~A} ~-~ {1 \over 2}
\hbar \psi^{~A}_{p} {\delta \over \delta e_{p}^{~AA'}} = 0~. \nonumber
\end{equation}
and its conjugate. At a Bianchi model, these constraints read:
\begin{eqnarray}
\bar S_{A'} \Psi & = & \epsilon^{pqr} e_{pA A'} \,\omega_{q~B}^{~A} \, \psi_r^{~B} \Psi ~-~ {1 \over 2}
\hbar \psi^{~A}_{p} {\partial \Psi \over \partial e_{p}^{~AA'}} = 0 \\
S_A \Psi & = & - \omega_{pA}^{~~B}  {\partial \Psi \over \partial \psi^{~B}_{p}}  ~+~ {1 \over 2} \,
 \,\hbar  D^{B A'}_{pq}  {\partial \over \partial \psi^{~B}_{p}}
 ~{\partial \Psi \over \partial e^{~A A'}_{q}} = 0~.
\end{eqnarray}
We notice that different $\omega_{pAB}$ correspond to different Bianchi class A models.
The Lorentz constraints \cite{84} are
\begin{eqnarray}
 J_{AB} & = & e_{p(B}^{~~~A'} {\partial \over \partial e_{P}^{~A)B'}} ~+~ 
\psi_{p(B}{\partial \over \partial \psi_p^{~A)}}, \\
 J_{A'B'} & = & e_{p~(B'}^{~A}{\partial \over \partial e_{P}^{|A|A')}}~. 
\end{eqnarray}
These two Lorentz constraints imply that the wave function should be invariant under the rotation in 
the spinor indices and depend on the three-geometry $h_{pq}$ of hypersurface only. So we can write
\begin{eqnarray}
\Psi & = & \phi_0(h_{mn}) ~+~ C_{pq}(h_{mn}) \, \psi^{pA}  \, \psi^q_A ~+~ V^{pqr}(h_{mn}) \,
n_{AA'} \, e_{pB}^{~~~A'} \, \psi_q^A \, \psi_r^B \nonumber \\
&  &  + \Psi_4 ~+~ \phi_6(h_{mn}) \prod^3_{i = 1} \psi^{iA} \, \psi_{iA}
\end{eqnarray}
where $C_{pq}$ is symmetric and $V^{pqr}$ is anti-symmetric in their last
two indices. The $C_{pq}$ and $V^{pqr}$ provide 6 and 9 degrees of freedom respectively. Also,
\begin{eqnarray*}
\Psi_4 &=& E_{1122} \psi^{1A} \psi_{1A} \psi^{2B} \psi_{2B} + E_{1133} \psi^{1A} \psi_{1A} \psi^{3B} \psi_{3B}
+ E_{2233} \psi^{2A} \psi_{2A} \psi^{3B} \psi_{3B} \\
&+& E_{1123} \psi^{1A} \psi_{1A} \psi^{2B} \psi_{3B} + E_{2213} \psi^{2A} \psi_{2A} \psi^{1B} \psi_{3B}
+ E_{3312} \psi^{3A} \psi_{3A} \psi^{1B} \psi_{2B} \\
&+& F^p_{~1233} e_{pB}^{~~A'} n_{AA'} \psi^{1A} \psi^{2B} \psi^{3C} \psi^3_{~C}
+  F^p_{~1323} e_{pB}^{~~A'} n_{AA'} \psi^{1A} \psi^{3B} \psi^{2C} \psi^2_{~C} \\
&+& F^p_{~2311} e_{pB}^{~~A'} n_{AA'} \psi^{2A} \psi^{3B} \psi^{1C} \psi^1_{~C}~. 
\end{eqnarray*}
The E's and V's also provide 6 and 9 degrees of freedom respectively.
These then give the most general solution to the Lorentz constraints. 
There is a duality relation between two fermions and four fermions \cite{84}.
By solving the two-fermion level, we can apply the Fourier transform \cite{84} to obtain the corresponding
four-fermion level. 

The above supersymmetry constraint and wave function are gauge invariant. We  use 
these gauge-invariant supersymmetry constraints to annihilate a gauge-invariant wave function and obtain all the 
equations of the theory. We then impose the condition of a diagonal Bianchi-IX metric in these equations. 
There is no loss of any physical information in the last step because the equations are 
derived from a gauge-invariant procedure.
The $\bar S_{A'}$ constraint is of first order, giving the derivatives ${\partial \Psi} / \partial
e_i^{~AA'}$, evaluated in particular at a Bianchi-IX model. Then the Bianchi-IX $S_A$ constraint
can be found as the hermitian adjoint of $\bar S_{A'}$. Indeed, the imposition of a diagonal Bianchi-IX metric isolates the 
true degrees of physical freedom because the isometry group of a Bianchi-IX universe is three dimensional. 
After completion of this work, we found that our Hamilton-Jacobi equation is equivalent to that derived in \cite{cg},
giving confirmation of our approach.

We also mention that because we impose the condition of a diagonal metric, there are no-off diagonal components
in our metric and hence there are no-off diagonal derivatives. If off-diagonal derivatives were present, we would lose some physical 
information in our equations and would not get the right Hamilton-Jacobi equation. To have derived the correct Hamilton-Jacobi 
equation in our case is a self-consistency check. Another justification is that we get the correct result for the bosonic order 
(see below Eqns.\ (13)-(16)) where the off-diagonal derivatives are not present.

The diagonal Bianchi-IX three-metric is given in terms of 
the three radii $A, B, C $ by
\begin{equation}
 h_{i j} = A^2 E^1_{~i} E^1_{~j} + B^2 E^2_{~i} E^2_{~j} + C^2 E^3_{~i}
E^3_{~j}
\end{equation}
where $ E^1_{~i}, E^2_{~i}, E^3_{~i} $ are a basis of unit left-invariant
one-forms on the three-sphere \cite{Ryan}. In the calculation, we shall repeatedly
need the expression, formed from the connection:
\begin{eqnarray}
\omega_{pAB} \, n^A_{~~B'} \, e^{qBB'} & = &  {i \over 4} \left( C^2 + B^2 - A^2 \right) 
\delta^1_p \, \delta^q_1 \nonumber\\
& +& {i \over 4} \left( A^2 + C^2 -B^2 \right) \delta^2_p \, \delta^q_2 \nonumber\\
& + & {i \over 4} \left( B^2 + A^2 -C^2 \right) \delta^3_p \, \delta^q_3~. \nonumber\\
\end{eqnarray}
We now  solve the supersymmetry constraints. First consider  $\bar S_{A'} \Psi = 0$
at the one-fermion level. One obtains
\begin{equation}
\epsilon^{pqr} \, e_{pAA'} \, \omega_{q~B}^{~A} \, \psi_r^B \, \phi_0 ~+~ 
\hbar \,  e_{qAA'} \, \psi_p^A {\partial \phi_0 \over \partial h_{pq}} = 0,
\end{equation}
where the relations $e_{pAA'} e_q^{~AA'} = - h_{pq}$ and $\partial / \partial 
e_p^{~AA'} = -2 e_{qAA'} \partial / \partial h_{pq}$ have been used.
 Since it is true for all $\psi_r^B$, one can conclude
\begin{equation}
\epsilon^{pqr} \, e_{pAA'} \, \omega_{q~B}^{~A} \, \phi_0 ~+~ \hbar \, e_{qBA'} \,
{\partial \phi_0 \over \partial h_{rq}} = 0~. 
\end{equation}
Multiply this equation by $e_l^{~BA'}$, giving
\begin{equation}
i \left( h_{ql} \, n^A_{~A'}  \, e^{pBA'} \omega_{pAB} ~-~ n^A_{~A'} e_q^{~AB'} \, \omega_{lAB} \right)
\, \phi_0 ~-~ \hbar \, h_{pl} h_{qs} {\partial \phi_0 \over \partial h_{ps} }= 0~.
\end{equation}
If $q \neq l$, this is an identity $ 0 = 0 $. Now take $q = l = 1$, say:
\begin{eqnarray*}
&  &  i A^2 \left[ {i \over 4} \left( C^2 + B^2 + A^2 \right) - {i \over 4} 
\left( C^2 + B^2 - A^2 \right) \right]  \phi_0 
 - \hbar \, h_{p1} \, h_{1s} \, {\partial \phi_0 \over \partial h_{pq}} = 0 \nonumber\\
 & \Rightarrow & \hbar {\partial \phi_0 \over \partial A } ~+~ A \, \phi_0 = 0~.
\end{eqnarray*}
Similarly for the $B, C$ dependence:
\begin{eqnarray}
\ &  & \hbar {\partial \phi_0 \over \partial B } ~+~ B \, \phi_0 = 0 \nonumber\\
\ &  & \hbar {\partial \phi_0 \over \partial C } ~+~ C \, \phi_0 = 0 \nonumber\\
\Rightarrow &  & \phi_{0} \propto \exp \left( -{1 \over 2 \hbar} \left( A^2 ~+~ B^2 ~+~ C^2 \right)
\right)~.
\end{eqnarray}
This is a well known result and has been first worked out in ref.\ \cite{P}.  Now let us study the  more
interesting two-fermion level. It turns out that the real physical degrees of freedom are provided
by $C_{11}, C_{22}, C_{33}, V^{123}, V^{231}, V^{312}$. The other coefficients of the two-fermion level
are zero and hence not physical. We will first derive the equations relating
$C_{12}, \cdots$ and $V^{112} \cdots$ and show that they are zero. 
After that, we will derive the equations for the physical amplitudes.

Consider $S_{A} \Psi = 0$ at the one-fermion level:
\begin{eqnarray}
& & 2 C_{pq} \, h^{qv}  \,\omega^p_{~AD} ~-~ 2 V^{tuv} \, \omega_{uA}^{~~B} \, n_{BC'} \,e_{tD}^{~~C'} 
\nonumber\\
&  & + 2 \hbar \, h^{tv} \, { \partial C_{pt} \over \partial h_{qr}} D_{D~~q}^{~A'p} \, e_{rAA'} 
~-~ 2 \hbar \, {\partial V^{tuv} \over \partial h_{pq}} D^{BA'}_{~~uq} \, e_{pAA'} \, n_{BC'} \, e_{tD}^{~~C'} 
\nonumber \\
&  & + \hbar \, V^{tuv} \, D^{BA'}_{~~uq} \, n_{AA'} \, e^q_{~BC'} \, e_{tD}^{~~C'}
~+~  i \hbar \, h_{ut} \, V^{tuv} \, \epsilon_{AD} = 0~.
\end{eqnarray}
Contracting the indices $A$ and $D$ with $\epsilon^{AD}$:
\begin{eqnarray}
\Rightarrow &  & - 2V^{tuv} \, \omega_{uA}^{~~B} \, n_{BC'} \, e_{t}^{~AC'} ~+~ i \hbar \,
{\partial V^{tuv} \over \partial h_{pq}} \, \left( 2h_{pt} \, h_{qu} ~-~ h_{pq} \, h_{ut} \right) \nonumber\\
&  & + {i \over 2} \,  \hbar \, h_{ut} \, V^{tuv} = 0~.
\end{eqnarray}
Notice that in the second term in Eqn.\ (18), all the indices of the metric components and metric derivatives are
contracted. Hence there is no loss of information when we impose the condition of a diagonal metric.
If one puts $v=1, 2, 3$, one obtains respectively,
\begin{eqnarray}
& & \hbar \left(B^3 {\partial \over \partial B} - AB^2 {\partial \over \partial A} -
 CB^2  {\partial \over \partial C}\right) V^{221} ~+~  \hbar \, B^2 \, V^{221} \nonumber\\
  & + &\hbar \left(C^3 {\partial \over \partial C} - AC^2 {\partial \over \partial A}
 - BC^2  {\partial \over \partial B}\right) V^{331}  ~+~ \hbar \, C^2 \, V^{331} \nonumber\\
 & + &B^2 \left( A^2 + C^2 - B^2\right) V^{221} ~+~ C^2 \left( A^2 + B^2 - C^2\right) V^{331} = 0~.
\end{eqnarray}
and two further equations which are just given by cyclic permutations of $A, B ,C$ and $1, 2, 3$.
We can also multiply (16) by $e_x^{~AD'} \, n^D_{~D'}$ and simplify the expressions. After some algebra, 
we obtain
\begin{eqnarray}
& & \hbar \left(2h_{xr} \, {\partial C_{pw} \over \partial h_{rp}} - h_{rs} \, 
{\partial C_{xw} \over \partial h_{rs}}\right) + 2 \,C_{pw}\, \omega^p_{~AD} 
\,n^D_{~D'} \, e_x^{~AD'} \nonumber\\
&-& \hbar\,{\partial V^{tuv} \over \partial h_{pq}} \, h_{qu} \,h_{vw} \, \epsilon_{txp} ~+~ {\hbar \over 2}
{\partial V^{tuv} \over \partial h_{pq}} \, h_{pq} \,h_{vw} \, \epsilon_{txu} \nonumber\\
&+& i \, h_{vw} \, V^{tuv} \, \epsilon_{txy} \, \omega_{uAB} \,  n^B_{~D'} \, e^{yAD'} 
~+~ {3 \over 4} \, \hbar \, h_{vw} \, V^{tuv} \, \epsilon_{txu} = 0~.
\end{eqnarray}
If we consider the off diagonal elements of $(x,w)$, we get six equations which are:

\begin{eqnarray}
& & {B^2 \over 4} \left[ \hbar ( C {\partial \over \partial C} - B {\partial \over \partial B} - 
A {\partial \over \partial A} )  + (A^2 + B^2 - C^2)  - 3 \hbar \right] V^{232} \nonumber\\
&+ & {i \over 2} \left[ \hbar (A {\partial \over \partial A} - B {\partial \over \partial B} - 
C {\partial \over \partial C}) ~+~  (C^2 + B^2 - A^2) \right] C_{12} = 0,
\end{eqnarray}
\begin{eqnarray}
& & {A^2 \over 4} \left[\hbar (C {\partial \over \partial C} - B {\partial \over \partial B} 
- A {\partial \over \partial A}) + (A^2 + B^2 - C^2)  -  3 \hbar \right] V^{113} \nonumber\\
& +& {i \over 2} \left[ \hbar (B {\partial \over \partial B} - A {\partial \over \partial A} 
- C {\partial \over \partial C}) + (C^2 + A^2 - B^2) \right] C_{12} = 0~.
\end{eqnarray}
The other four equations are also cyclic permutations of the above on $A, B ,C$ and $1, 2, 3$.
Now we consider $\bar S_{A'} \Psi = 0$ at three-fermion level. Following  similar
methods, we get the following equations from $\psi_1 \psi_2 \psi_2$:
\begin{eqnarray}
& \hbar& \left[ (h^{22})^2 \, {\partial C_{22} \over \partial h_{11}} \, e_{1BA'}  ~-~
  h^{11} \, h^{22} \, {\partial C_{12}  \over \partial h_{22}} \, e_{2BA'} \right] \nonumber \\
& + &\left[(h^{22})^2 \,C_{22} \, \epsilon^{pq1} ~-~ h^{11} \, h^{22} \,C_{12} \, \epsilon^{pq2} 
\right] e_{pAA'} \, \omega_{a~B}^{~A} \nonumber \\
& + & \left(\hbar \,{\partial V^{s21} \over \partial h_{22}} e_{2~A'}^{~C} 
~+~ \epsilon^{pq2} \, V^{s21} \, e_{pAA'} \, \omega_q^{~AC}\right) n_{CC'} \, e_{sB}^{~~C'} \nonumber\\
& +& {\hbar \over4} \, V^{s21} \left(n_{CA'} e^2_{~BC'} e_s^{~CC'} - n^C_{~A'} e^2_{CC'} e_{sB}^{~~C'} \right)
~+~ {3\over4} \, \hbar V^{221} \,n_{BA'} = 0.
\end{eqnarray}
Multiplying Eq.\ (22) by $n^{BA'}$, we obtain the result
\begin{displaymath}
V^{221} = f_{221}(A,C) \, {1 \over B^3} \, e^{- {B^2 \over 2 \hbar}},
\end{displaymath}
where $f_{221}(A,C)$ is an arbitrary function of $A, C$. We can also multiply (22) by $e_l^{~BA'}$. In these cases
for $l=2$, $l=3$ respectively, we obtain
\begin{eqnarray*}
 C_{12} & = & f_{12}(A,C) \, e^{- {B^2 \over 2 \hbar}}, \\
 V^{121} & = & f_{121}(A,C) \, {1 \over B} \, e^{- {B^2 \over 2 \hbar}}.
\end{eqnarray*}
Similarly, considering $\psi_1 \psi_3 \psi_3,~ \psi_2 \psi_1 \psi_1, \cdots$, we have
\begin{eqnarray}
& &C_{12} = f_{12}(C) \, e^{- {{A^2 + B^2} \over 2 \hbar}},~
C_{13} = f_{13}(B) \, e^{- {{A^2 + C^2} \over 2 \hbar}},~ 
C_{23} = f_{23}(A) \, e^{- {{B^2 + C^2} \over 2 \hbar}}, \nonumber\\
& &V^{112} = f_{112}(C) \, {1 \over A^3B} \, e^{- {{A^2 + B^2} \over 2 \hbar}},~
V^{113} = f_{113}(B) \, {1 \over A^3C} \, e^{- {{A^2 + C^2} \over 2 \hbar}}, \nonumber\\
& &V^{212} = f_{212}(C) \, {1 \over AB^3} \, e^{- {{A^2 + B^2} \over 2 \hbar}},~
V^{223} = f_{223}(A) \, {1 \over B^3C} \, e^{- {{B^2 + C^2} \over 2 \hbar}}, \nonumber\\
& &V^{313} = f_{313}(B) \, {1 \over AC^3} \, e^{- {{A^2 + C^2} \over 2 \hbar}},~
V^{323} = f_{323}(A) \, {1 \over BC^3} \, e^{- {{B^2 + C^2} \over 2 \hbar}}. 
\end{eqnarray}
If we consider $\bar S_{A'} \Psi = 0$ at the order corresponding to $ \psi_1^E \psi_2^F \psi_3^G $,
 we  have three free unprimed indices 
$E, F, G$ and one free primed index $A'$. We can contract two of the three unprimed indices, say $F,G$, to get
\begin{eqnarray}
&  & ~\left(4h^{22}h^{33}\epsilon^{pq1}C_{23} - 2h^{11}h^{33}\epsilon^{pq2}C_{13}
- 2h^{11}h^{22}\epsilon^{pq3}C_{12}\right) e_{pAA'} \omega_{q~E}^{~A} \nonumber\\
&-& 2\left( V^{S13} \epsilon^{pq2} ~+~ V^{s12} \epsilon^{pq3}\right) \omega_q^{~AF}n_{EC'}e_{sF}^{~~C'}\nonumber\\
&+& \hbar \left(4h^{22}h^{33} {\partial C_{23} \over \partial h_{11} } e_{1EA'}
- 2h^{11}h^{33} {\partial C_{13} \over \partial h_{22} } e_{2EA'}
- 2h^{11}h^{22} {\partial C_{12} \over \partial h_{33} } e_{3EA'}\right)\nonumber\\
&+& 2 \hbar \left({\partial V^{s12} \over \partial h_{33}} e_{3GA'} n_{EC'} e_s^{~GC'}
- {\partial V^{s13} \over \partial h_{22}} e_{2~A'}^{~F} n_{EC'} e_{sF}^{~~C'}\right)\nonumber\\
&+& {1\over2} \hbar \, V^{s13}\, n^F_{~A'} \,(e^2_{~EC'} \, e_{sF}^{~~C'} + e^2_{~FC'} \, e_{sE}^{~~C'})\nonumber\\
&+& {1\over2} \hbar \,V^{s12}\, n^F_{~A'}\, (e^3_{~EC'}\,e_{sF}^{~~C'} + e^3_{~FC'}\,e_{sE}^{~~C'})\nonumber\\
&-& {3 \over 2} \hbar\, \left( V^{213} ~+~ V^{312} \right) n_{EA'} = 0.
\end{eqnarray}
Multiplying the last expression by $e_l^{~EA'}$ for $l = 1, 2, 3$ will give
\begin{eqnarray}
&& {2 \over B^2 C^2} \left(\hbar A {\partial C_{23} \over \partial A} + A^2 \, C_{23}\right)
+ {i \over2} \left({\hbar \over B} {\partial V^{313} \over \partial B} + {\hbar \over B^2} \,V^{313}
+ V^{313}\right) \nonumber\\
&& - {i \over2} \left({\hbar \over C} {\partial V^{212} \over \partial C} + {\hbar \over C^2} \, V^{212}
+ V^{212}\right) = 0,
\end{eqnarray}
\begin{equation}
{1 \over A^2 C^2} \left( \hbar B {\partial C_{13} \over \partial B} + B^2\, C_{13}\right)
- {i \over2} \left( {\hbar \over C} {\partial V^{112} \over \partial C} + {\hbar \over C^2}
V^{112} + V^{112} \right) = 0,
\end{equation}
\begin{equation}
{1 \over A^2 B^2} \left( \hbar C {\partial C_{12} \over \partial C} + C^2\, C_{12}\right)
+ {i \over2} \left( {\hbar \over B} {\partial V^{113} \over \partial B} + {\hbar \over B^2}
V^{113} + V^{113} \right) = 0~.
\end{equation}
with their cyclic permutations. Using (25), (26), (27) and their cyclic permutations,
one can easily prove that the only solutions satisfying these equations are
\begin{eqnarray}
& & C_{12} \propto  e^{-{1\over2 \hbar}(A^2 +B^2 + C^2)}, ~ 
C_{13} \propto e^{-{1\over2 \hbar}(A^2 +B^2 + C^2)} , ~
C_{23} \propto  e^{-{1\over2 \hbar}(A^2 +B^2 + C^2)}, \nonumber\\
& & V^{112} \propto {1 \over A^3BC} e^{-{1\over2 \hbar}(A^2 +B^2 + C^2)},~ 
V^{113} \propto {1 \over A^3BC} e^{-{1\over2 \hbar}(A^2 +B^2 + C^2)}, \nonumber\\
& &V^{212} \propto {1 \over AB^3C} e^{-{1\over2 \hbar}(A^2 +B^2 + C^2)},~ 
V^{223} \propto {1 \over AB^3C} e^{-{1\over2 \hbar}(A^2 +B^2 + C^2)}, \nonumber\\
& & V^{313} \propto {1 \over ABC^3} e^{-{1\over2 \hbar}(A^2 +B^2 + C^2)}, ~
V^{323} \propto {1 \over ABC^3} e^{-{1\over2 \hbar}(A^2 +B^2 + C^2)}.
\end{eqnarray}
However, if we substitute the above 9 amplitudes back into (18), (20) and (21) and their cyclic permutations, they will not
satisfy the equations. Hence the only solutions are zero. We can see that these 9 amplitudes are not
the dynamical degrees of freedom of the theory, which are contained in the remaining 6 coefficients.
Below we  will derive the rest of the
equations for the remaining coefficients at two-fermion level. It will be verified in the next section how these
equations  provide the dynamics of the theory.

We now come back to $S_A \Psi = 0$ at one-fermion level and $\bar S_{A'} \Psi = 0$ at three-fermion level. We have only
considered the off-diagonal elements of $(x,w)$ of (19). From the diagonal elements, one obtains
 coupled partial differential equations between the physical amplitudes.
\begin{eqnarray}
& & A^2 \left[ {\hbar \over 2} \, \left( C {\partial \over \partial C} - A {\partial \over \partial A}
- B {\partial \over \partial B}\right) + \left(A^2 + B^2 - C^2\right) -{3\over2} \hbar \right] 
 V^{231} \nonumber\\
&+ & A^2 \left[ {\hbar \over 2} \, \left( C {\partial \over \partial C} + A {\partial \over \partial A}
- B {\partial \over \partial B}\right) - \left(A^2 + C^2 - B^2\right)  + {3\over2} \hbar \right] 
 V^{321} \nonumber\\
&+& i \hbar \left( A {\partial \over \partial A} - C {\partial \over \partial C}
- B {\partial \over \partial B}\right)C_{11} + i \left( C^2 + B^2 -A^2 \right)C_{11} = 0,
\end{eqnarray}
\begin{eqnarray}
& & B^2 \left[ {\hbar \over 2} \, \left( - C {\partial \over \partial C} + A {\partial \over \partial A}
+ B {\partial \over \partial B}\right) - \left(A^2 + B^2 - C^2\right)  + {3\over2} \hbar \right] 
 V^{132} \nonumber\\
&+ & B^2 \left[ {\hbar \over 2} \, \left(- C {\partial \over \partial C} + A {\partial \over \partial A}
- B {\partial \over \partial B}\right) + \left(B^2 + C^2 - A^2\right)  - {3\over2} \hbar \right] 
 V^{312} \nonumber\\
&+& i \hbar \left( B {\partial \over \partial B} - C {\partial \over \partial C}
- A {\partial \over \partial A}\right)C_{22} + i \left( C^2 + A^2 -B^2 \right)C_{22} = 0,
\end{eqnarray}

\begin{eqnarray}
& &  C^2 \left[ {\hbar \over 2} \, \left( C {\partial \over \partial C} - A {\partial \over \partial A}
+ B {\partial \over \partial B}\right) - \left(C^2 + B^2 - A^2\right) + {3\over2} \hbar \right] 
 V^{213} \nonumber\\
&+ & C^2 \left[ {\hbar \over 2} \, \left(- C {\partial \over \partial C} - A {\partial \over \partial A}
+B {\partial \over \partial B}\right) + \left(A^2 + C^2 - B^2\right)  - {3\over2} \hbar \right] 
 V^{123} \nonumber\\
&+& i \hbar \left( - A {\partial \over \partial A} - C {\partial \over \partial C}
+ B {\partial \over \partial B}\right)C_{33} + i \left( A^2 + B^2 -C^2 \right)C_{33} = 0.
\end{eqnarray}
But if we multiply (22) by $e_1^{~BA'}$, we have
\begin{equation}
\hbar A {\partial C_{22} \over \partial A} + A^2 C_{22}
- {i \over 2} B^2 \left(\hbar B {\partial \over \partial B} + \hbar + B^2\right)  V^{321} = 0,
\end{equation}
and the cyclic permutations
\begin{equation}
\hbar B {\partial C_{33} \over \partial B} + B^2 C_{33}
- {i \over 2} C^2 \left(\hbar C {\partial \over \partial C} + \hbar + C^2\right)  V^{132} = 0,
\end{equation}
\begin{equation}
\hbar C {\partial C_{11} \over \partial C} + C^2 C_{11}
- {i \over 2} A^2 \left(\hbar A {\partial \over \partial A} + \hbar + A^2\right)  V^{213} = 0.
\end{equation}
Also multiply the equation obtained from $\psi_1 \psi_3 \psi_3$ by  $e_1^{~BA'}$, we have
with the cyclic permutations
\begin{equation}
\hbar B {\partial C_{11} \over \partial B} + B^2 C_{11}
+ {i \over 2} A^2 \left(\hbar A {\partial \over \partial A} + \hbar + A^2\right)  V^{312} = 0,
\end{equation}
\begin{equation}
\hbar C {\partial C_{22} \over \partial C} + C^2 C_{22}
+ {i \over 2} B^2 \left(\hbar B {\partial \over \partial B} + \hbar + B^2\right) V^{123} = 0,
\end{equation}
\begin{equation}
\hbar A {\partial C_{33} \over \partial A} + A^2 C_{33}
+ {i \over 2} C^2 \left(\hbar C {\partial \over \partial C} + \hbar + C^2\right)  V^{231} = 0.
\end{equation}
We obtain three more equations by multiplying Eq.\ (24) by $n^{EA'}$, and taking cyclic permutations:
\begin{equation}
\left( \hbar B {\partial \over \partial B} + B^2 + 3 \hbar\right) V^{213}
+ \left( \hbar C {\partial \over \partial C} + C^2 + 3 \hbar\right) V^{312} = 0,
\end{equation}
\begin{equation}
\left( \hbar A {\partial \over \partial A} + A^2 + 3 \hbar\right) V^{123}
+ \left( \hbar C {\partial \over \partial C} + C^2 + 3 \hbar\right) V^{321} = 0,
\end{equation}
\begin{equation}
\left( \hbar A {\partial \over \partial A} + A^2 + 3 \hbar\right) V^{132}
+ \left( \hbar B {\partial \over \partial B} + B^2 + 3 \hbar\right) V^{231} = 0.
\end{equation}

We have found all the equations relating $C_{11}, C_{22}, C_{33}$ and $V^{123}, V^{231}, V^{312}$.
In the next section, we are going to investigate the semi-classical solution of the above equations.

\section{\bf Semi-Classical Solutions to Supersymmetry Constraints}

We have seen  there are twelve equations for $C_{11}, C_{22}, C_{33}, V^{123}, V^{231}, V^{312}$, 
namely (29)-(40). Here we use these equations to make a comparsion with Csord\'as and Graham \cite{cg},
who found that a  Hartle-Hawking state \cite{haw} exists semi-classically.
Here we check that  our system of first order-partial differential equations also admits a Hartle-Hawking
state semi-classically by studying the Hamilton-Jacobi equation. We assume that the coefficients have the form
\begin{eqnarray}
C_{11} & = & \left( C_{(0)11} + \hbar C_{(1)11} + \hbar^2 C_{(2)11} + \cdots\right) e^{-I/ \hbar},
~{\rm etc.}\ , \nonumber\\
V^{123} & = &  \left( V_{(0)}^{123} + \hbar V_{(1)}^{123} + \hbar^2 V_{(2)}^{123} 
+ \cdots\right) e^{-I/ \hbar}, ~{\rm etc.,} 
\end{eqnarray}
where $I$ is a classical Euclidiean action.

Consider the Hamilton-Jacobi equation. Substituting (41) into (29)-(40), and collecting all the terms of 
order $\hbar^0$, we have
\begin{eqnarray}
& & i \left( -A{\partial I \over \partial A} + B{\partial I \over \partial B} 
+ C{\partial I \over \partial C} - A^2 - B^2 - C^2 \right) C_{(0)11} \nonumber\\
& +& {A^2 \over 2}  \left( B {\partial I \over \partial B} + A{\partial I \over \partial A} 
- C{\partial I \over \partial C} + A^2 + B^2 - C^2 \right) V_{(0)}^{231} \nonumber\\
& +&  {A^2 \over 2}  \left( B {\partial I \over \partial B} - A{\partial I \over \partial A} 
- C{\partial I \over \partial C}  - A^2 + B^2 - C^2 \right) V_{(0)}^{321} = 0,
\end{eqnarray}
\begin{equation}
C\left( -{\partial I \over \partial C} + C \right) C_{(0)11} +
 {i \over2} A^3 \left(-{\partial I \over \partial A} + A \right) V_{(0)}^{231} = 0,
\end{equation}
\begin{equation}
B \left( -{\partial I \over \partial B} + B \right) C_{(0)11} -
 {i \over2} A^3 \left(-{\partial I \over \partial A} + A \right) V_{(0)}^{321} = 0,
\end{equation}
\begin{equation}
 B\left( -{\partial I \over \partial B} + B \right) V_{(0)}^{231} +
C\left( -{\partial I \over \partial C} + C \right) V_{(0)}^{321} = 0.
\end{equation}
Equations (43), (44) and (45) are consistent with each other. From (43), (44), we subsitute
 the $ V$'s into (42) to get an equation homogenous in $ C_{(0)11}$ only. To have a non-trivial solution of
$C_{(0)11}$, the coefficent must be zero, giving
\begin{eqnarray}
& & A^2 \left( {\partial I \over \partial A}\right)^2 + B^2 \left( {\partial I \over \partial B}\right)^2 
+ C^2 \left( {\partial I \over \partial C}\right)^2 
- 2AB \left( {\partial I \over \partial A}\right) \left( {\partial I \over \partial B}\right)\nonumber\\
&-& 2AC \left( {\partial I \over \partial A}\right) \left( {\partial I \over \partial C}\right)
-2BC \left( {\partial I \over \partial B}\right) \left( {\partial I \over \partial C}\right) \nonumber\\
&-& A^4 - B^4 - C^4  + 2A^2B^2 + 2A^2C^2 + 2B^2C^2 = 0.
\end{eqnarray}
For the other $C$'s and $V$'s, one  obtain the same Hamilton-Jacobi equation\footnote{It can be checked
that the Hamilton-Jacobi equation is equivalent to the one that can be derived in \cite{cg}}. 
From dimensional ground, the action $I$ has the general form,
\begin{equation}
I= \alpha \, A^2 + \beta \, B^2 + \gamma \, C^2 + \mu \, AB + \nu \, BC + \, \lambda \, AC~.
\end{equation}
Substituting this action $I$ into above Hamilton-Jacobi equation, it gives
\begin{eqnarray}
&&~(4 \alpha^2 -1) \, A^4 + (4 \beta^2 -1) \, B^4 + (4 \gamma^2 -1) \, C^4 \nonumber\\
&&+(2-8 \alpha \beta) \, A^2 B^2 + (2-8 \gamma \alpha) \, A^2 C^2 + (2-8 \beta \gamma) \, B^2 C^2 \nonumber\\
&&-4(2 \alpha \nu + \mu \lambda) A^2BC - 4 (2 \beta \gamma + \nu \mu) AB^2 C 
-4 (2 \gamma \mu + \nu \lambda) ABC^2 = 0~. \nonumber\\
\end{eqnarray}
Since the polynomials are independent of each other, each coefficient vanishes identically.
Hence the most general  solutions are
\begin{eqnarray}
\pm I &=& {1\over2} \left( A^2 + B^2 + C^2 \right) \nonumber\\
\pm I &=& {1\over2} \left( A^2 + B^2 + C^2 \right) - AB  - AC - BC \nonumber\\
\pm I &=& {1\over2} \left( A^2 + B^2 + C^2 \right) + AB  + AC - BC \nonumber\\
\pm I &=& {1\over2} \left( A^2 + B^2 + C^2 \right) + AB  - AC + BC \nonumber\\
\pm I &=& {1\over2} \left (A^2 + B^2 + C^2 \right) - AB + AC + BC .
\end{eqnarray}
The first one is the wormhole action and the second one is the Hartle-Hawking action. 
At least in the semi-classical level, both wormhole and Hartle-Hawking state exist in the 
same fermion sector. Our results in here exactly correspond to  Csord\'as and Graham \cite{cg}. 

\section{Conclusion And Discussion}

In section 2 we carried out the dimensional reduction and obtained the supersymmetry constraints;
this involves writing down the most general solution to the Lorentz constraints. We then solved for the 
supersymmetry constraints and found that 9 out of 15 degrees of freedom at the two-fermion level are not physical.
The coupled first order partial differential equations describing the remaining 6 degrees of freedom  were given 
in section 3. We solved the Hamilton-Jacobi equation completely and found the complete set of solutions.
Both the Hartle-Hawking and wormhole actions are among the solutions. 

We also mention that the Ansatz of the wave function constructed by  Csord\'as and Graham \cite{cg} may only 
work if there are  no chiral breaking terms in the supersymmetry constraints as in pure $N = 1$ supergravity.
 Supersymmetry constraints
with no chiral breaking terms will preserve the number of fermions. The presence of  chiral breaking terms will not 
conserve the number of fermions and gives mixing of  different levels of fermions. This occurs (e.g.) when 
$N = 1$ supergravity is coupled to supermatter \cite{WB}. However, our approach 
can readily be generalized to non-chiral models. 

In the future, we hope to study  inhomogenous perturbations of a Friedmann $k= +1$ model in supersymmetric
quantum cosmology, using spectral  boundary conditions for gravitinos \cite{hal} (Bianchi IX models
are a particular kind of distortion of a k = +1 model). It will be interesting to see if a 
Hartle-Hawking state still exists in these models.

\section*{\bf Acknowlegements}

A.D.Y.C. thanks the Croucher Foundation of Hong Kong for financial support.


\begin{thebibliography}{99}

\bibitem{finite}  P.~D.~D'Eath, Supersymmetric Quantum Cosmology (Cambridge University Press 1996), in press

\bibitem{pilati} M.\ Pilati, Nucl.\ Phys.\ {\bf B132}, 138 (1978)

\bibitem{84} P.~D.~D'Eath, Phys.~Rev.~{\bf D29}, 2199 (1984)


\bibitem{P} P.~D.~D'Eath, Phys.\ Rev.\ {\bf D48}, 713 (1993)

\bibitem{P1} P.~D.~D'Eath, S.~W.~Hawking and O.~Obreg\'on, Phys.\ Lett.\ {\bf 300}B, 44 (1993)

\bibitem{P2} R.~Capovilla and J.\ Guven, Class.\ Quant.\ Grav.\ {\bf 11}, 1961 (1994)

\bibitem{gl}  R.\ Graham and H.\ Luckock,  Phys.\ Rev.\ {\bf D49}, 4981 (1994)
 
\bibitem{PP} A.~D.~Y.~Cheng, P.~D.\ D'Eath and P.~R.~L.~V.~Moniz,  Phys.~Rev.~{\bf D49}, 5246 (1994);
R.~Capovilla and O.~Obreg\'on, Phys.\ Rev.\ {\bf D49}, 6562 (1994)

\bibitem{cg} A.~Csord\'as and R.~Graham, Phys.\ Rev.\ Lett.\ {\bf 74}, 4926 (1995)

\bibitem{T} C.~Teitelboim, Phys.\ Rev.\ Lett.\ {\bf 30}, 1106 (1977)

\bibitem{WB} J.~Wess and J.~Bagger, Supersymmetry and Supergravity 2nd edition (Princeton University
Press, Princeton 1992)

\bibitem{Ryan} M.~P.~Ryan and L.~L.~Shepley, Homogeneous Relativistic Cosmologies (Princeton University
Press, Princeton 1975)

\bibitem{haw} J.~B.~Hartle and S.~W.~Hawking, Phys.~ Rev.~{\bf D28}, 2960 (1983)

\bibitem{worm} S.~W.~Hawking, Phys.~ Rev.~{\bf D37}, 904 (1988)

\bibitem{hal}  P.~D.~D'Eath and J.~J.~Halliwell, Phys.~ Rev.~{\bf D35}, 1100 (1987)

\end{thebibliography}
\end{document}